\documentclass[preprint,showpacs,preprintnumbers,amsmath,amssymb]{revtex4}
\usepackage{graphicx}
\usepackage{amsmath}
\draft

\begin{document}
\title{Thermodynamic black di-rings}
\author{Hideo Iguchi and Takashi Mishima} 
\affiliation{
Laboratory of Physics,~College of Science and Technology,~
Nihon University,\\ Narashinodai,~Funabashi,~Chiba 274-8501,~Japan
}
\date{\today}
\begin{abstract}
Previously the five dimensional $S^1$-rotating black rings have been superposed in a concentric way by some solitonic methods, and regular systems of two $S^1$-rotating black rings were constructed by the authors and then Evslin and Krishnan (we called these solutions ``black di-rings").
In this place we show some characteristics of the solutions of five dimensional black di-rings, especially in thermodynamic equilibrium.
After the summary of the di-ring expressions and their physical quantities, first we comment on the equivalence of the two different solution sets of the black di-rings. Then the existence of thermodynamic black di-rings is shown, in which both isothermality and isorotation between the inner black ring and the outer black ring are realized. 
We also give detailed analysis of peculiar properties of the thermodynamic black di-ring 
including discussion about a certain kind of thermodynamic stability (instability) of the system.
\end{abstract}      
\maketitle

\section{Introduction}

Since the two epoch-making discoveries of Myers-Perry black holes \cite{Myers:1986un} and $S^1$-rotating black rings by Emparan and Reall \cite{Emparan:2001wn},  several nontrivial black hole systems have been obtained and used to clarify peculiarities of the higher dimensional gravity.
Especially in five dimensions, solitonic 
solution-generation methods invented in the legendary
era from the 1970s to the 1980s have been proved to be
greatly powerful and applied to systematic reconstruction of known solutions including the
Myers-Perry black holes  \cite{Pomeransky:2005sj} and the $S^1$-rotating black rings
\cite{Iguchi:2006rd,Tomizawa:2006vp}, and also to construction of new solutions--the $S^2$-rotating black
rings \cite{Mishima:2005id,Iguchi:2006tu,Tomizawa:2005wv} and doubly rotating black rings \cite{Pomeransky:2006bd}, for example.
As a further trial, hunting new solutions has been ambitiously attempted to obtain novel black holes that have never been expected in four dimensions, like black hole systems with multihorizons \cite{Elvang:2007rd,Iguchi:2007is,Evslin:2007fv,Izumi:2007qx,Elvang:2007hs} or a topologically nontrivial horizon \cite{Evslin:2008gx,Chen:2008fa}.

%
One of the main reasons to collect various types of higher dimensional black holes is to accomplish the phase structure of higher dimensional black objects and classify the properties into the universal ones independent of the dimensional number of spacetimes and the peculiar ones depending on the dimensional number. 
To perform the systematic analyses the information deduced from the solutions (if possible all solutions)  corresponding to each phase is necessary, so that finding new solutions is required. 
So far for the cases of six and higher dimensional asymptotically Minkowski spacetimes, the analyses have mainly depended on the metric form of Myers-Perry black holes and the perturbed metric about the solutions \cite{Dias:2009iu,Dias:2010eu,Dias:2010maa}, or  the approximate black ring solutions based on the matching between the metrics of a thin ring and a black string \cite{Emparan:2009vd}. 
As a promising recent advance, numeric methods have also been introduced to decide the stability of black objects, though limited to topologically spherical black holes \cite{Shibata:2010wz}.

%
Up to the present the solitonic methods have not yet been available to higher than five dimensions with globally asymptotic Minkowski because of the mathematical structure of the methods. 
So if we proceed over the case of a single horizon and extensively study the phase structure of multihorizon systems, the five dimensional solutions that can be systematically obtained are useful though the five dimensions are rather marginal dimensionality in comparison with six and higher dimensions.
%
%
Actually as pioneering works to argue the phase of multihorizon systems \cite{Elvang:2007hg}, the black Saturns have been already investigated and were shown to have thermodynamic equilibrium states, and furthermore, some clues to clarify the phase structures of higher dimensional cases were extracted (see also \cite{Emparan:2007wm}).
On the other hand, so far there is not any analysis of  multihorizon systems using the black di-rings.
Mainly we may think the following two reasons for this fact.
First, the authors of the works mentioned above suggested nonexistence of thermodynamic equilibrium plausibly  \cite{Elvang:2007hg}.
Second, the expressions of black di-rings are considerably intricate. 
In this paper we therefore show some properties of a black di-ring, particularly that of the di-ring in thermodynamic equilibrium (hereafter called ``thermodynamic black di-ring"). 

%
Black di-rings are concentric configurations composed of two independently $S^1$-rotating black rings.    
The authors first discovered the regular di-rings by using the solitonic method similar to the B\"{a}cklund transformation  (this is called ``di-ring I" hereafter) \cite{Iguchi:2007is}. 
Successively, Evslin and Krishnan constructed another di-ring solution set (called ``di-ring II") \cite{Evslin:2007fv}. 
They used the inverse scattering method that was modified by Pomeransky to treat the higher dimensional case (hereafter abbreviated to PISM) \cite{Pomeransky:2005sj}. 
However, because of the complexity of their expressions, the attempt to confirm the equivalence of these two 
di-ring solution sets and further investigation of the physics of the di-ring systems still remain to be done. 
So we precede the study with giving a confirmation of equivalence between the two solution sets mentioned above.

%
For the systems of the black di-rings concerned here, we consider five dimensional spacetimes with three commuting Killing vector fields: a timelike Killing vector field and two axial-Killing vector fields. We assume further that one of the axial-Killing vector fields is orthogonal to the other. So the line elements adopted here can be reduced to 
\begin{equation}
\label{eq:eq1}
ds^2 =G_{t t}(dt)^2+2G_{t \psi}dt d\psi +G_{\psi \psi}(d\psi)^2 
      +G_{\phi\phi}(d\phi)^2 
        \,+\,e^{2\nu}\left(d\rho^2+dz^2\right)\,,
\end{equation}
where the metric coefficients are the functions of $(\rho,z)$ and $\det G = -\rho^2$ is imposed.
It should be noticed that owing to the assumptions any type of angular momentum corresponding to $\phi$ rotation, Arnowitt-Deser-Misner (ADM)  angular momentum and Komar angular momentum corresponding to each ring vanishes.
That is, only the cases with a single rotation are considered. 
At the present even the study of the cases with a single rotation is far from completed, so that some clues may be derived through the analysis given this place. 

%
The rest of this paper is as follows. 
In Sec. II we show the complete equivalence of these different solution sets of di-rings, after the representations and the physical quantities of both the black di-ring I and II are presented. 
The confirmation is done with the aid of the facts established by Hollands and Yazadjiev \cite{Hollands:2007aj}, which concern the uniqueness of higher dimensional black holes.
This equivalence ensures that one can use a more convenient representation between the black di-ring I and II.
In Sec. III we first describe the conditions for a regular black di-ring to become a thermodynamic black di-ring. 
%
%
The conditions for the thermality together with the regularity conditions for elimination of conical singularities are solved in substance and reduced to a remarkably simple form to perform the analysis. 
Using the reduced constraints we also obtain the simply reduced physical variables of the thermodynamic black di-rings as described in Appendix \ref{sec:variable3}. 
%
%
In the rest of the section, using these quantities the existence of thermodynamic black di-rings and some interesting properties of the systems are shown.
Section IV is devoted to the analysis of whether thermodynamic stabilities (local stabilities) are realized or not in the system of the thermodynamic di-ring.
In Sec. V concluding remarks are given. 

\section{Representations of black di-rings}

We consider briefly where the difference between the black di-ring I and II comes from and how to verify the equivalence of the two  di-rings, and also give the necessary quantities for the study. 
The solutions of the black di-rings (i.e., di-ring I and II) have been constructed by rather different solitonic solution-generation techniques, respectively, so that the expressions of the solutions are considerably different. 
Furthermore, the solitonic methods generally cause redundant complexity on the derived solutions. 
It therefore becomes a nontrivial (rather formidable) task to determine whether the two di-ring systems are equivalent or not with straightforward algebraic calculation.
Hence we must use some powerful mathematical facts to do it, as mentioned later. 
Here we show that there is an onto and one-to-one mapping between them and give just a rough sketch of the procedure to verify it.
%
%
A fully detailed explanation will be given elsewhere.
%
\begin{figure}
\centering
\noindent
\includegraphics[keepaspectratio=true,width=11cm]{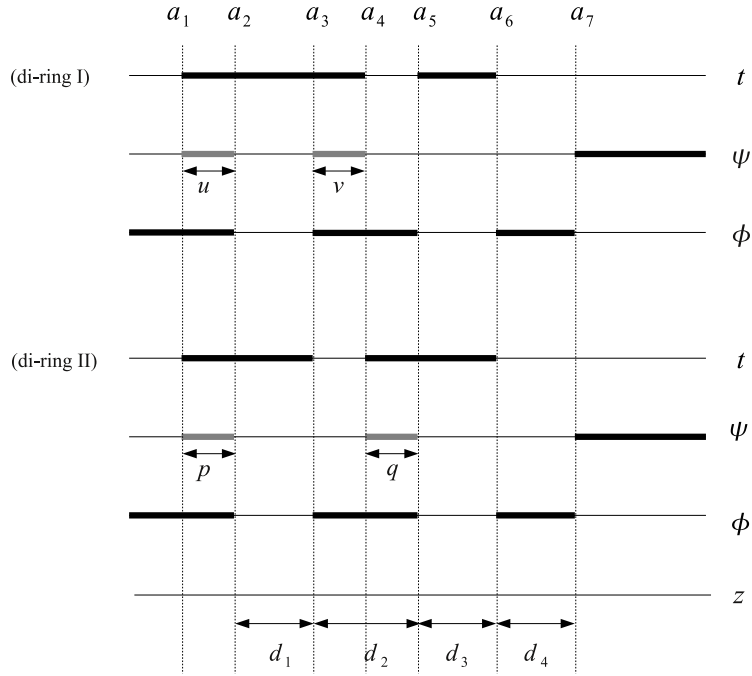}
\noindent
\caption{
Rod structures describing the seeds of the di-rings. 
The upper and lower rod diagrams correspond to the di-ring I and di-ring II, respectively. 
Black rods are assigned to $1/2$ line mass density, while gray rods (holes) to $-1/2$ line mass density. 
Two solitons are removed and recovered at the positions $a_1$ and $a_4$ for the di-ring cases. 
}
\label{fig:II1}
\end{figure}
%
%

%
It is helpful to view the both di-ring systems from the same standpoint.
So it is noteworthy that the solutions of di-ring I can be also generated by PISM from appropriate seeds as the di-ring II. 
This fact can be verified mainly using the following two facts: 
(1) as verified in Ref. \cite{Tomizawa:2006jz} any two-soliton solution generated by the method of B\"{a}cklund transformation that was first adopted in Ref. \cite{Mishima:2005id} generally corresponds to a two-soliton solution obtained by the original inverse scattering method that was established by Belinski and Zakharov \cite{Belinsky:1971nt,Belinsky:1979mh};
%
%
(2) in the single-rotational case we can easily transform the procedure of the original inverse scattering method into  that of PISM. 
This means that both the di-ring systems can be viewed from the standpoint of the PISM. 
%

%
The new solutions are generated by adding the appropriate solitons to the seed solutions, following the procedure of PISM: (i) removing solitons with trivial Belinski-Zakharov (BZ) parameters, (ii) scaling appropriately the metric obtained in  process (i), (iii) constructing a generating matrix and recovering the same solitons as in step (i), but with nontrivial BZ parameters, (iv) scaling back for process (ii), and (v) filling ``flaws'' which remain on the axes using the arbitrariness of BZ parameters. 
The seeds to be used to generate black di-ring I and II are depicted as rod structures \cite{Harmark:2004rm} in Fig. \ref{fig:II1}. 
We assign $d_{i}\,(i=1,2,3,4)$ to the length of the $i$th finite rod and also the rod itself. 
The parameters $(u,v)$ of the di-ring I and $(p,q)$ of the di-ring II inscribed on Fig.1 are associated with lengths of the gray rods (holes) and have the following range: 
\begin{eqnarray}
\label{eq:II1}
&&0\leq u \,, \ \ 0\leq v \leq d_2\,,\ \\
\label{eq:II2}
&&0\leq p \,, \ \ 0\leq q \leq d_2\,.
\end{eqnarray}
For both the di-rings, in the above step (i) two solitons with trivial BZ parameters are removed at the positions $a_1$ and $a_4$; then in step (iii) the same solitons with nontrivial BZ parameters are put back at the same positions. 
Here $(b^{\rm (I)}_{\rm L},\, b^{\rm (I)}_{\rm R})$ and $(b^{\rm (II)}_{\rm L},\, b^{\rm (II)}_{\rm R})$ are assigned  to BZ parameters of di-ring I and II, respectively, whose subscript ${\rm L}$ (${\rm R}$) means that the quantity is associated  with the soliton at $a_{1}$ ($a_{4}$).
The BZ parameters are fixed through process (v) as follows 
\footnote{Note that there is ambiguity of the generating matrix in step (iii). The BZ parameters which cure the singular behavior on the axes depend on the choice of the generating matrix. In the solution generations, we use the generating matrices to which the BZ parameters (4) and (5) correspond.},
\begin{eqnarray}
\label{eq:II3}
&&\left\{\,
 b^{\rm (I)}_{\rm L} =\pm \left( \frac{2a_{21} a_{61} a_{71}}{a_{31} a_{51}} \right)^{1/2}\,, \  
 b^{\rm (I)}_{\rm R} =\pm \left( \frac{2a_{42} a_{64} a_{74}}{a_{43} a_{54}} \right)^{1/2}\,
  \,\right\}
  \\
\label{eq:II4}
&&\left\{\,
 b^{\rm (II)}_{\rm L} =\pm \left( \frac{2a_{31} a_{61} a_{71}}{a_{21} a_{51}} \right)^{1/2}\,, \  
 b^{\rm (II)}_{\rm R} =\pm \left( \frac{2a_{43} a_{64} a_{74}}{a_{42} a_{54}} \right)^{1/2}\,
\,\right\}\,,
\end{eqnarray}
where the symbols $a_{ij}$ are defined as $a_{i}-a_{j}$. 
As consequences of these processes, we obtain the regular di-ring systems except conical singularities.  
The sets of moduli parameters $\{u,v,d_1,d_2,d_3,d_4 \}_{\rm I}$ and $\{p,q,d_1,d_2,d_3,d_4 \}_{\rm II}$ provide the solution sets of di-ring I and II, respectively. 
Rods $d_{1}$ and $d_{3}$ correspond to outer and inner black rings, respectively, while rods $d_{2}$ and $d_{4}$  correspond to outer and inner axes of $\phi$ rotation.
Hereafter the subscripts L and R are also assigned to ``outer'' and ``inner,'' respectively, according to the positions of the rods.  
If we want to eliminate conical singularities that generally occur on the outer $\phi$ axis ($d_{2}$) and the inner $\phi$ axis ($d_{4}$), we must impose two constraints on each set of the moduli parameters. 
%

%
It is clear from the preceding paragraph that the essential difference between the black di-ring I and II comes from the difference of the corresponding seeds.
Particularly positions of two ``holes'' give an essential difference, because the black di-ring I and II do not match generally even if all rods $d_{i}$ and the positions of two solitons ($a_{1}$ and $a_{4}$) are the same in the black di-ring I and II like Fig. \ref{fig:II1}.   
Each hole controls the rotation of the corresponding ring. 
Generally these two moduli sets have a complicated relationship except the case that the holes corresponding to $v$ and $q$ vanish. 
%

%
Once the metric functions of the di-rings are written down, physical quantities can be computed from the metric in principle. 
The physical quantities for the black di-ring I and II are collected in Appendixes A and B, while the full expressions of the metric functions will be given elsewhere. 
From the expressions in the appendixes, we notice that the physical quantities of the di-ring I and II can be easily written down with only the moduli set  $\{u,v,d_1,d_2,d_3,d_4 \}_{\rm I}$ or $\{p,q,d_1,d_2,d_3,d_4 \}_{\rm II}$, respectively. 
%
This fact means that all the physical quantities are expressed with only $a_{ij}$.
The following expressions of ADM masses ($M^{\rm (I)}$ and $M^{\rm (II)}$), periodic angles ($\Delta\phi_{\rm L}^{\rm (I)}$ and $\Delta\phi_{\rm L}^{\rm (II)}$) needed to keep regularity on the $\phi$ axis $d_2$  ,
and other periodic angles ($\Delta\phi_{\rm R}^{\rm (I)}$ and $\Delta\phi_{\rm R}^{\rm (II)}$) needed to keep regularity on the $\phi$ axis  $d_4$ 
are useful for the next study, 
\begin{eqnarray}
\label{eq:II5}
M^{\rm (I)}&=&
\frac{3\pi}{4} \left( a_{65} + a_{41} \frac{b^{{\rm (I)}\,2}_{\rm R} - b^{{\rm (I)}\,2}_{\rm L}+2 a_{41}}
                          {(b^{\rm (I)}_{\rm R}-b^{\rm (I)}_{\rm L})^2} \right)\,, 
\\ 
\label{eq:II5_2}
M^{\rm (II)}&=&
\frac{3\pi}{4}\left( a_{31}+a_{64} \right) ,  
\\
\label{eq:II6}
\left( \frac{\Delta\phi_{\rm L}^{\rm (I)}}{2\pi} \right)^{2}&=& 
\frac{a_{53}a_{62}a_{73}^{2}(a_{42}a_{31}b^{\rm (I)}_{\rm L} -  a_{21}a_{43}b^{\rm (I)}_{\rm R} )^2 }
    {  a_{21}a_{31}a_{42}a_{43}a_{52}a_{63}a_{72}^2(b^{\rm (I)}_{\rm L}-b^{\rm (I)}_{\rm R})^2  }\,, \ \ \ \ \ \ 
\\
\label{eq:II7}
\left( \frac{\Delta\phi_{\rm L}^{\rm (II)}}{2\pi} \right)^{2}&=& 
\frac{a_{21}a_{42}a_{53}a_{62}a_{73}}{2a_{31}a_{43}}
\left(
   \frac{a_{43}a_{51}b^{\rm (II)}_{\rm L} +  a_{31}a_{54}b^{\rm (II)}_{\rm R} }
        {a_{41}a_{52}a_{63}a_{72}}
\right)^2\,,
\\
\label{eq:II8}
\left( \frac{\Delta\phi_{\rm R}^{\rm (I)}}{2\pi} \right)^{2}&=&
\frac{a_{73}^{2}a_{76}
 ( a_{74}b^{\rm (I)}_{\rm L} -  a_{71}b^{\rm (I)}_{\rm R} )^2 }
    {  a_{71}a_{72}^{2}a_{74}a_{75}(b^{\rm (I)}_{\rm L}-b^{\rm (I)}_{\rm R})^2  }\,, 
\\
\label{eq:II8_2}
\left( \frac{\Delta\phi_{\rm R}^{\rm (II)}}{2\pi} \right)^{2}&=&
 \frac{a_{71} a_{74} a_{73} a_{76}}{a_{72}^2 a_{75}^2}\,,
\end{eqnarray}
where Eqs. (\ref{eq:II3}) and (\ref{eq:II4}) are used to describe the BZ parameters with moduli parameters.  
%

%
We shall give a brief explanation to conclude the equivalence of the di-ring I and II. 
A key mathematical fact to establish the equivalence of the di-ring systems is in the work by Hollands and Yazadjiev,  who have discussed the uniqueness of five dimensional stationary black holes with asymptotic flatness and axial $U(1)^2$ symmetry \cite{Hollands:2007aj} (see also \cite{Morisawa:2004tc} and \cite{Morisawa:2007di}).
Originally they have considered the systems of a single black hole, but their proof can be applied to the systems of multiple black holes so that their theorem is still valid with some modification \cite{Armas:2009dd}.  
From the mathematical fact deduced from their works, we can infer that two different systems of black di-ring become  isometric when all the lengths of rods and the Komar angular momenta coincide with each other.
The Komar angular momenta can be replaced with two other independent physical quantities, but some auxiliary conditions are usually needed to resolve the discrete degeneracy remaining (see the example below).    
In the case of the di-rings, using the lengths of rods in place of the interval structures is enough for the  mathematical fact to be valid, because the rod vectors are trivial. 
Furthermore, it should be noticed that the procedure of the proof seems to be independent of the existence of conical singularities on the axes. 
So the above statement is generalized to the case with conical singularities.
Thus we can conclude that two di-rings are isometrically equivalent if the following equality of the ADM masses and the periodic angles are satisfied with some auxiliary condition: 
\begin{eqnarray}
\label{eq:II9}
&&\{d_1,d_2,d_3,d_4 \}^{\rm (I)}=\{d_1,d_2,d_3,d_4 \}^{\rm (II)}   \\
\label{eq:II10}
&&M^{\rm (I)}= M^{\rm (II)}\,,\ \ \  \Delta\phi^{\rm (I)}_{\rm R}=\Delta\phi^{\rm (II)}_{\rm R}. 
\end{eqnarray}
Here the choice of the appropriate signs of $b^{\rm (I)}_{\rm L}\,b^{\rm (I)}_{\rm R}$ and $b^{\rm (II)}_{\rm L}\,b^{\rm (II)}_{\rm R}$ plays a role in the  auxiliary condition.
%

%
As a final step, we shall give a mathematical relationship between the sets of black di-ring I and II. 
When the condition (\ref{eq:II9}) is satisfied, the remaining condition (\ref{eq:II10}) can be considered as the simultaneous equations to 
provide a mapping function between $(u,v)$ space and $(p,q)$ space:
\begin{equation}
\label{eq:II11}
\left\{\begin{array}{ll}
    &M^{\rm (I)}(u,v)=M^{\rm (II)}(p,q)      \\
    &\Delta\phi^{\rm (I)}_{\rm R}(u,v)=\Delta\phi^{\rm (II)}_{\rm R}(p,q).
\end{array}\right. 
\end{equation}
%
\begin{figure}
\centering
\vskip -1.7cm
\noindent
\includegraphics[keepaspectratio=true,width=12.5cm]{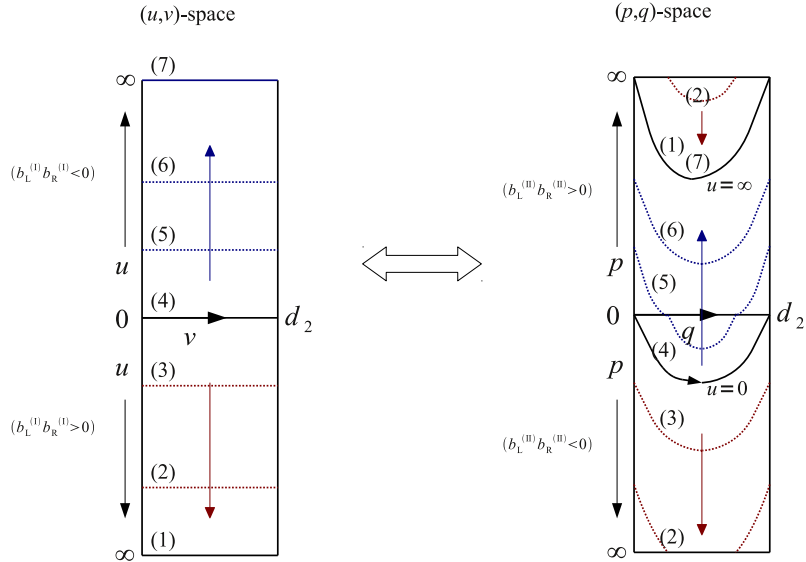}
\vskip -0.5cm
\noindent
\caption{
Correspondence between the space of $(u,v)$ for di-ring I and the space of $(p,q)$ for di-ring II.
Each horizontal line with a number from (1) to (7)  is mapped into the curve with the same number. 
}
\label{fig:II2}
\end{figure}

At this point, if we notice from the expressions (\ref{eq:II5_2}) and (\ref{eq:II8_2}) that the ADM mass and regular periodic angle of the  di-ring~II have remarkably simple forms with respect to $p$ and $q$, Eq. (\ref{eq:II11}) can be solved analytically and  give an explicit mapping function [i.e., $(p,q)=(p(u,v),q(u,v))$]. 
Examining the mapping function, we can confirm that the correspondence between $(u,v)$ space and $(p,q)$ space is onto and one-to-one, if infinities are included in these parameter spaces.
Thus we can say that the systems of black di-ring I and II are completely equivalent. 
In Fig. \ref{fig:II2} the correspondence between $(u,v)$ space and $(p,q)$ space is schematically shown. 
After the equivalence is established, we can choose a more convenient representation among di-ring I and di-ring II to investigate the physics of the di-ring according to the problem we face. For example, the representation of  di-ring I is more suitable for the analysis of the case that the angular velocity of $\psi$ rotation of the inner ring is zero, while  di-ring II is more suitable for the case that the Komar angular momentum of $\psi$ rotation of the inner ring is zero. 
%
%

\section{Thermodynamic black di-ring}

The existence of regular black di-rings has been already confirmed in Refs. \citep{Iguchi:2007is} and Ref. \citep{Evslin:2007fv}.
Here we show the existence of thermodynamic black di-rings (i.e., systems of regular black di-ring in thermodynamic equilibrium) and their peculiar properties \footnote{According to Emapran's comment at YKIS 2010, similar results seem to be found by Emparan and Figueras.}.
The physical quantities used here are normalized with the ADM mass $M$ to compare the physical quantities of black objects following \cite{Elvang:2007rd}.
The normalized quantities corresponding to ADM angular momentum, area of horizon, temperature of horizon, and angular velocity of horizon are given as follows:   
\begin{equation}
\label{eq:III1}
j^2=\frac{27\pi}{32G}\frac{J^2}{M^3}\,,\ \,\,
a_{h} = \frac{3}{16}\sqrt{\frac{3}{\pi}} \frac{A_{h}}{(GM)^{3/2}}\,,\ \,\,
\tau_{h} = \sqrt{\frac{32\pi}{3}} (GM)^{1/2} T_{h}\,,\ \,\,
\omega_{h}=\sqrt{\frac{8}{3\pi}} (GM)^{1/2} \Omega_{h}\,.
\end{equation}
To ensure that a system of regular black di-ring becomes a thermodynamic equilibrium state, 
the temperatures ($\tau_{\rm L},\,\tau_{\rm R}$) and angular velocities ($\omega_{\rm L},\,\omega_{\rm R}$ ) of the outer (L) and inner (R) black rings of the system must satisfy the following condition,  
\begin{equation}
\label{eq:III2}
\tau_{\rm L}=\tau_{\rm R}\,,\ \ \ \omega_{\rm L}=\omega_{\rm R}\,.
\end{equation}
It should be noticed that compared with the case of black Saturn, to construct a thermodynamic black di-ring with two black rings is not so trivial as pointed out in \cite{Elvang:2007hg}.  
Actually black rings with the same temperature and angular velocity have the same shape and size \cite{Elvang:2006dd}, so that we must arrange two identical rings into a thermodynamic di-ring keeping the thermality above mentioned and also the balance between gravitational attraction and centrifugal forces.  
That is, it seems rather natural that thermodynamic di-rings do not appear in the phase space. 
If there remains any possibility of thermodynamic di-rings, some nonlinearity may play a key role.
%

%
First we show the existence of thermodynamic equilibrium of a regular black di-ring by solving the above constraints  together with the following balance conditions (geometrically to eliminate conical singularities on the axes),  
\begin{equation}
\label{eq:III3}
\left( \frac{\Delta\phi_{\rm L}}{2\pi} \right)^2=1\,,\ \ \ \left( \frac{\Delta\phi_{\rm R}}{2\pi} \right)^2=1\,.
\end{equation}
For convenience, we introduce other moduli parameters composed of the rods $d_{i}$ $(i=1 \sim 4)$ as follows: 
\begin{equation}
\label{eq:III4}
\begin{array}{lll}
 h_1 = d_{1}+d_{2}+d_{3}+d_{4}\,(= a_{72})\,,    &\ \ h_3 = d_{3}+d_{4}\,(= a_{75})\,,          \\
 h_2 = d_{2}+d_{3}+d_{4}\,(= a_{73})\,,          &\ \ h_4 = d_{4}\,(= a_{76})\,.  
\end{array}
\end{equation}
Fortunately the conditions (\ref{eq:III2}) and (\ref{eq:III3}) can be nearly solved in an analytic way, though rather lengthy calculation is needed.
In fact the four equations in (\ref{eq:III2}) and (\ref{eq:III3}) are reduced to 
\begin{eqnarray}
\label{eq:III5}
0&=&(4 h_{1}^2 - 7 h_{1} h_{4} + 4 h_{4}^2)\,h_{3}^3
 - (8 h_{1}^3 - 8 h_{1}^2 h_{4} - h_{1} h_{4}^2 + 4 h_{4}^3)\,h_{3}^2  \nonumber   \\
&& + (5 h_{1}^4 - 2 h_{1}^3 h_{4} - 2 h_{1}^2 h_{4}^2 + h_{1} h_{4}^3 + h_{4}^4)\,h_{3}-h_{1}^5\ \ \ \ \ \ \ \ \ \ {\rm (for\ I\ and\ II)}\,,  \\
\label{eq:III6}
h_{2}&=&h_{1}-h_{3}+h_{4}\qquad\qquad\qquad\qquad\qquad\qquad\ \ \ \ \ \ \ \ \ \ \ \ \ \ \ {\rm (for\ I\ and\ II)}
\end{eqnarray}
and also 
\begin{equation}
\label{eq:III7}
u-v =\frac{(h_{1}-h_{3})(h_{2}^2 + 2h_{1}h_{3} - h_{3}^2) }{h_{3}h_{4}}\,,\ \ \ 
(u+h_{1})(h_{2}-v) =\frac{h_{1}^2h_{2}^2 }{h_{3}h_{4}}\ \ \ \ {\rm (for\ I)}
\end{equation}
or 
\begin{equation}
\label{eq:III8}
p+q =\frac{(h_{1}-h_{2})(h_{3}^2 + 2h_{1}h_{2} - h_{2}^2) }{h_{2}h_{4}}\,,\ \ \ 
(p+h_{1})(q+h_{3}) =\frac{h_{1}^2h_{3}^2 }{h_{2}h_{4}}\ \ \ \ {\rm (for\ II)}\,.
\end{equation}
The procedure to determine the moduli parameters using these equations is the following: 
(i) one of the parameters is fixed using the arbitrariness of global scaling--for example, $h_{1}$ is fixed to 1; 
(ii) we choose $h_{4}$ as a free parameter; (iii) candidates $(h_{2},h_{3})$ are given by solving Eq. (\ref{eq:III5}) and then using Eq. (\ref{eq:III6}); (iv) we select the set satisfying $h_{1}=1>h_{2}>h_{3}>h_{4}>0$; (v) from (\ref{eq:III7}) or (\ref{eq:III8}), we compute $(u,v)$ or $(p,q)$. 
It is noteworthy that the most nontrivial equation (\ref{eq:III5}) is cubic, so that the solutions can be treated analytically in principle.  
Furthermore, from Eq. (\ref{eq:III7}) or (\ref{eq:III8}), we can eliminate the hole-parameters $(u,v)$ or $(p,q)$ in the physical quantities, so that the expression of physical quantities of the thermodynamic black di-ring is remarkably simplified. 
The collection of the reduced physical variables of the thermodynamic black di-ring is given in Appendix C. 
%

\begin{figure}[]
\centering
\includegraphics[keepaspectratio=true,width=12cm]{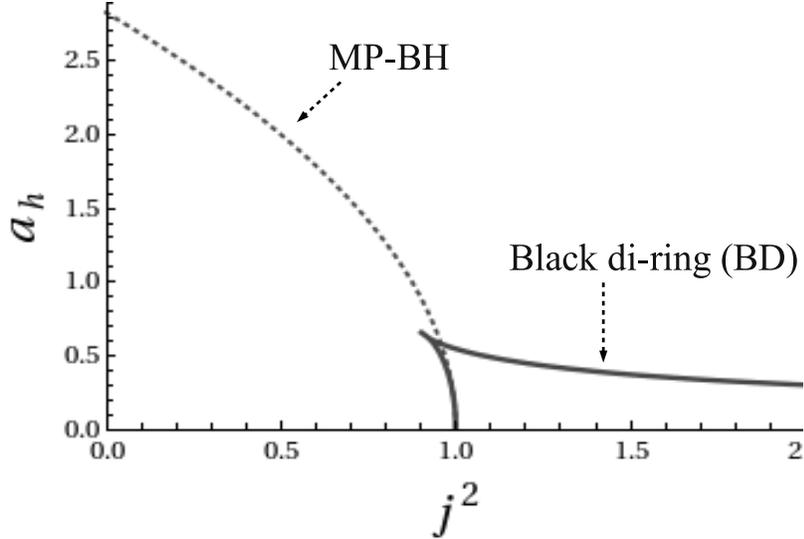}
\caption{
Total area $a_{h}$ vs angular momentum $j^2$. The phase of the thermodynamic black di-ring is described with a continuous curve (thick). A dotted curve shows states of the MP black hole. 
} 
\label{fig:III_1}
\end{figure}

%
Following the above procedure we can easily show the existence of the thermodynamic black di-ring. 
In Fig. \ref{fig:III_1} the states of the thermodynamic black di-ring appear as a continuous curve with a cusp. 
The minimum angular momentum ($j^{2} \approx 0.920~75$) and the maximum total area ($a_{h} \approx 0.622~54$) occur at the cusp.   
The cusp also is considered as the boundary where the behavior of the thermodynamic black di-ring is qualitatively divided into two branches like the black ring. 
So the branch where the state of the di-ring approaches the extremum of the MP black hole (i.e., $j^2=1$) is called ``fat ring'' or ``fat'' and the other is ``thin ring'' or ``thin''. 
%

\begin{figure}[]
\centering
\includegraphics[keepaspectratio=true,width=11.5cm]{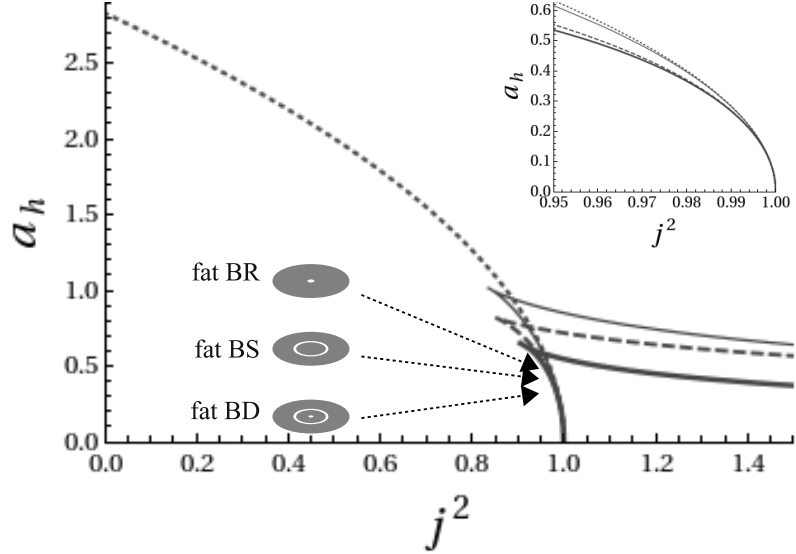}
\caption{
Total area $a_h$ as a function of angular momentum $j^2$.    
Dotted, thin, dashed, and thick curves correspond to MP black hole (MP-BH), black ring (BR), 
black Saturn (BS) and black di-ring (BD) respectively. 
}
\label{fig:III_2}
\end{figure}

\begin{figure}[]
\centering
\includegraphics[keepaspectratio=true,width=12.5cm]{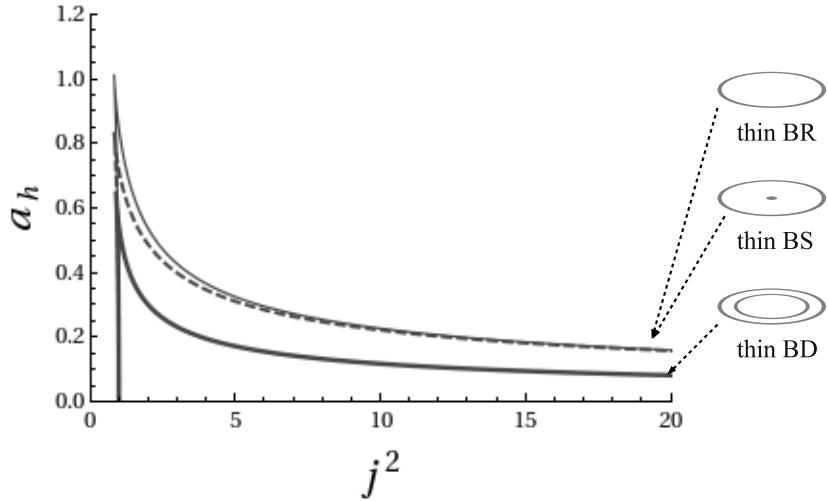}
\caption{
Total area $a_h$ as a function of angular momentum $j^2$. The behavior of the black objects in the thin branch is shown. 
}
\label{fig:III_3}
\end{figure}

%
The behavior of several thermodynamic black objects including the di-ring is shown in Figs. \ref{fig:III_2} and \ref{fig:III_3}.
The thick line in the figures shows the states of the thermodynamical black di-ring (BD). 
Other thermodynamic systems--Myers-Perry black hole (MP-BH), black ring (BR) and black Saturn (BS)--are also shown to be  compared with the di-ring. 
The state of the di-ring is less dominant than that of others with respect to entropy (i.e., area $a_h$) in both branches. 
In the fat branch, the curves of the black ring, black Saturn, and black di-ring approach that of the MP black hole, respectively, as depicted in Fig. \ref{fig:III_2}. 
More detailed behavior of the four black objects is shown in the upper right diagram magnified near the extremal point. 
An interesting fact is that each pair (MP black hole and black ring, black Saturn and black di-ring) has similar behavior near the extremal point.
%

\begin{figure}[]
\centering
\vskip -0.9cm
\includegraphics[keepaspectratio=true,width=12.5cm]{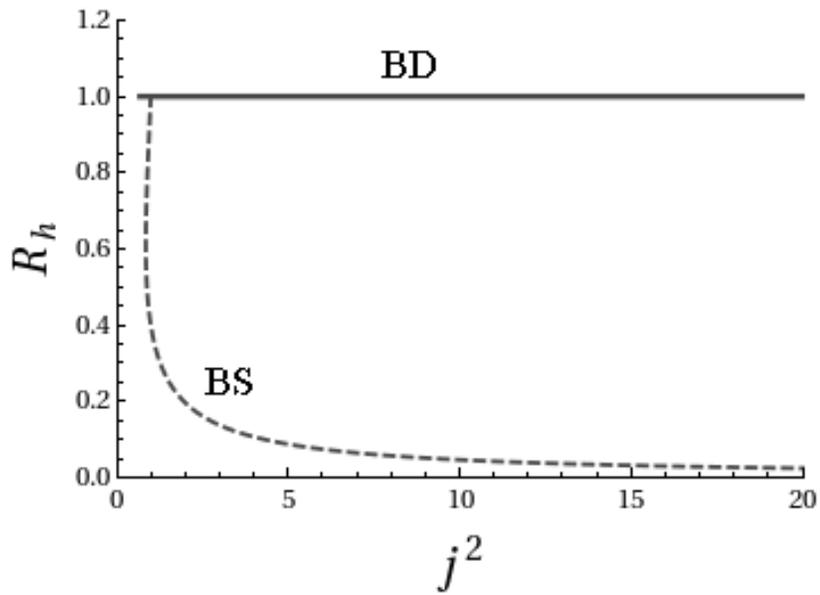}
\vskip -0.6cm
\caption{
Ratio of horizon area of the central black hole or inner black ring to that of the outer black ring ($R_{h}$). 
The dashed curve and thick curve are for black Saturn (BS) and black di-ring (BD), respectively.  
}
\label{fig:III_4}
\end{figure}

%
In the thin branch, from the right part of Fig. \ref{fig:III_3} the curve of black Saturn immediately asymptotes to that of the black ring while the black di-ring acts independently.  
Actually we can confirm that as $j^2$ increases infinitely, while the area of black Saturn approaches that of the black ring,  that of the black di-ring approaches a half of the black ring.  
This behavior can be explained by considering the configurations of the thin thermodynamic black objects. 
For the case of black Saturn, as already explained in the Ref. \cite{Elvang:2007hg}, in order to keep the isothermality, the longer the outer ring is, the larger the fraction of area and mass of the black Saturn the outer ring carries. 
In other words, in the thin limit the influence of the central black hole becomes neglected. 
For the case of the black di-ring, on the contrary, the inner ring must always stay with the outer ring like twins to keep the isothermality. 
So we can expect the behavior of the thin black di-ring to be qualitatively different from that of the thin black Saturn. 
Figure \ref{fig:III_4} also contrasts the black Saturn and black di-ring clearly, which shows the behavior of the ratio of the horizon area of the inner black ring or central black hole to the horizon area of the outer black ring ($R_{h}$). 
The ratio of the di-ring holds a constant value one, and that of the black Saturn quickly vanishes as the angular momentum $j^2$ increases. 
As a remarkable peculiarity of the di-ring, it is noteworthy that the inner ring and outer ring of the black di-ring always have the same area of horizon (i.e., entropy). 
The latter fact can be analytically confirmed from the reduced constraint (\ref{eq:III6}). 
This equation means $d_{1}=d_{3}$ (i.e., $a_{32}=a_{65}$), and hence from the equations from (\ref{sec:variable2}9) to (\ref{sec:variable2}12) the equality  between the areas of the inner and outer rings is verified owing to the isothermality. 

\clearpage
\section{Thermodynamic stability (instability)}

The black di-ring in thermodynamic equilibrium is not thermodynamically stable because the entropy of it is not the maximum. 
However, there is a possibility that the thermodynamic equilibrium is a local maximum of the entropy. 
Evslin and Krishnan \cite{Evslin:2008py} found the metastability of black Saturn which occurs when the reduced angular momentum lies in a narrow window 
$0.854~83 < j^2 < 0.854~94$ of the thin ring branch.
In this section we investigate whether the thermodynamic equilibrium of black di-ring is metastable or not.
Here we use the representations of black di-ring II.

The problem we consider is constrained extremization $a_h$ under the condition that $j^2$ is fixed. There are also additional constraints 
$\Delta \phi_{\rm L}=2\pi$ and $\Delta \phi_{\rm R}=2\pi$. The number of the parameters is six, 
$h_1$, $h_2$, $h_3$, $h_4$, $x$ and $y$ where $x = p + q + h_1 + h_3$ and $y = (p + h_1)(q + h_3)$.
Using the scaling freedom we can fix one of the parameters, e.g. $h_1=1$. The balance condition $\Delta \phi_{\rm R}=2\pi$ can be easily solved for $y$. Here we eliminate a parameter $y$ from the equations.  As a result, the physical variables are functions of four parameters. In the following  we abstractly denote the four parameters as $x_i$ where $i=1,\dots, 4$, for example, $a_h=a_h(x_1,x_2,x_3,x_4)$.
Now the problem is the extreme value problem of the function 
\begin{equation}
 f(x_1,x_2) = a_h(x_1,x_2,\varphi_3(x_1,x_2),\varphi_4(x_1,x_2))
\end{equation}
where $x_3=\varphi_3(x_1,x_2)$ and $x_4=\varphi_4(x_1,x_2)$ are obtained by formally solving the constraints 
\begin{equation}
\label{eq:const1}
 g_1(x_1,x_2,x_3,x_4) = j^2(x_1,x_2,x_3,x_4)-j_0^2=0
\end{equation}
and 
\begin{equation}
\label{eq:const2}
 g_2(x_1,x_2,x_3,x_4) = \Delta \phi_R(x_1,x_2,x_3,x_4)-2\pi=0
\end{equation}
for $x_3$ and $x_4$.

\begin{figure}
  \includegraphics[scale=0.8,angle=0]{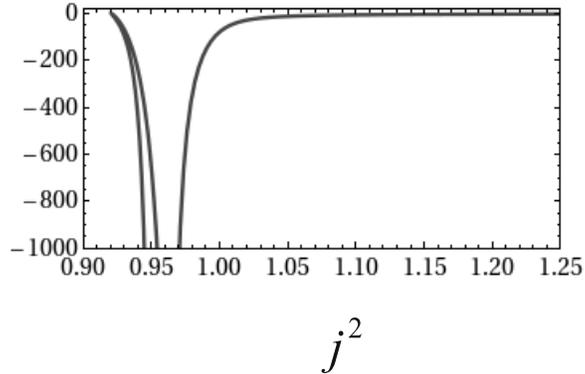}
  \caption{Eigenvalue of $H_{ij}$ (eigenvalue 1) vs reduced angular momentum $j^2$. The eigenvalue 1 is negative except when $j^2$ is in $0.920~75 < j^2 < 0.920~84$ of thin ring branch.}
 \label{fig:eigen1}
\end{figure}

\begin{figure}
  \includegraphics[scale=0.8,angle=0]{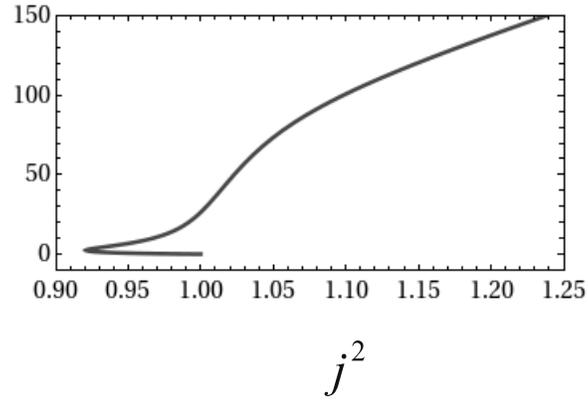}
  \caption{Eigenvalue of $H_{ij}$ (eigenvalue 2) vs reduced angular momentum $j^2$. The eigenvalue 2 is positive for all black di-rings in thermodynamic equilibrium.}
 \label{fig:eigen2}
\end{figure}

\begin{figure}
  \includegraphics[scale=0.8,angle=0]{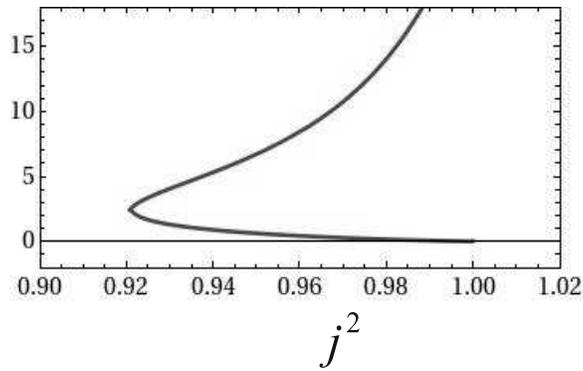}
  \caption{Close-up of the eigenvalue 2. We can see that the eigenvalue 2 is positive near the extremal limit.}
 \label{fig:eigen2_up}
\end{figure}

\begin{figure}
  \includegraphics[scale=0.8,angle=0]{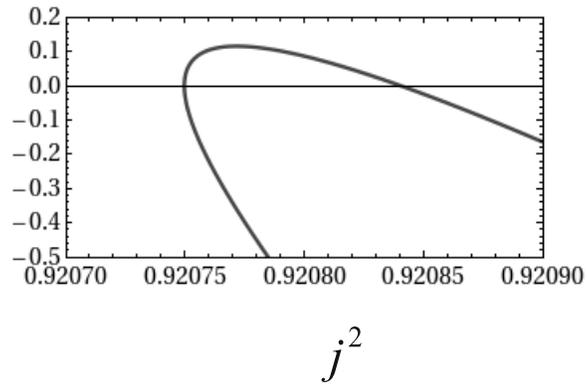}
  \caption{Close-up of the eigenvalue 1. The eigenvalue 1 becomes positive when $j^2$ is in $0.920~75 < j^2 < 0.920~84$ of the thin ring branch.}
 \label{fig:eigen1_up}
\end{figure}

It can be numerically confirmed that the area of the thermodynamic black di-ring is extremal in the moduli space where the first partial derivatives of $f$ with respect to $x_1$ and $x_2$ vanish.
The thermodynamic equilibria are metastable when the $2 \times 2$ matrix
\begin{equation}
 H_{ij} \equiv \frac{\partial^2 f(x_1,x_2)}{\partial x_i \partial x_j}
\end{equation}
has only negative eigenvalues. In the analysis we need the partial derivatives of $\varphi_3(x_1,x_2)$ and $\varphi_4(x_1,x_2)$. They are obtained by simultaneously solving the partial derivatives of the constraints (\ref{eq:const1}) and  (\ref{eq:const2}).
We numerically plot the eigenvalues of $H_{ij}$ as a function of $j^2$ for the thermodynamic equilibrium black di-rings in Fig. \ref{fig:eigen1} and \ref{fig:eigen2} where we set $x_1=h_2$, $x_2=h_4$, $x_3=h_3$ and $x_4=x$. It seems that one of the eigenvalues is negative (eigenvalue 1) while the other is positive (eigenvalue 2). Indeed, the eigenvalue 2 is always positive as in Fig. \ref{fig:eigen2_up}. Therefore the window of the metastability is closed. Zooming in Fig. \ref{fig:eigen1} as Fig. \ref{fig:eigen1_up}, we can find a narrow window $0.920~75 < j^2 < 0.920~84$ of the thin ring branch where both eigenvalues are positive. In this region the entropy of the thermodynamic equilibria is locally minimum.  It is likely that the interaction between the black rings makes the system thermodynamically unstable.


\clearpage
\section{Summary and Discussion}
%
In this paper the study has been devoted to clarifying several properties of the five dimensional thermodynamic black di-ring as a sequel to the finding of the regular black di-ring. 
%

%
First we verified the equivalence of the two different solution sets of the black di-ring I and II with the aid of the mathematical fact established by Hollands and Yazadjiev. 
This mathematical fact is powerful to judge whether two solutions are physically identical, but this does not mean that we can inform what kind of correspondence two solution sets have. 
So we constructed explicitly the mapping function from Eq. (\ref{eq:II11}) to complete the verification of the equivalence. 
%

%
Next we showed the existence of the thermodynamic black di-ring (regular black di-ring in thermodynamic equilibrium).  
We found that the conditions (\ref{eq:III2}) and (\ref{eq:III3}) to ensure the existence of thermodynamic black di-rings are reduced to the unexpectedly simple equations from (\ref{eq:III5}) to (\ref{eq:III8}). 
Using the reduced equations we confirmed the existence, and also clarified some intriguing properties of the phase of the thermodynamic black di-ring. 
The phase of the di-ring is described with a continuous curve with a cusp, which divides the curve into the fat ring branch and the thin ring branch. 
In the fat branch, as $j^2 \to 1$ the curve of the di-ring asymptotes to that of the MP black hole as the curves  of the black ring and the black Saturn do. 
And further, by magnifying the behavior of the four black objects near the extremal point it seems that the four  objects are paired off: the MP black hole and the black ring; the black Saturn and the di-ring.
In the thin branch, as the angular momentum increases, while the black Saturn promptly asymptotes to the black ring, the black di-ring rather behaves as another different branch. 
Actually we found that in the thin limit $j^2 \to \infty$, the ratio of the total area of the black di-ring to that of the black ring becomes one-half. 
When we go further to the analysis of more general thermodynamic multiobjects (tri-ring, black Saturn with more rings, etc.), it may be an interesting problem to what extent these results are generalized.   
In addition, it may be possible to extend the analysis to the unbalanced case \cite{Herdeiro:2009vd,Herdeiro:2010aq,Astefanesei:2009mc}.

%
Incidentally, we found that the inner ring and outer ring of the thermodynamic black di-ring always have the same horizon area (i.e., entropy). 
To construct the thermodynamic black di-ring two identical black rings must be prepared, so that these black rings are symmetric for permutation before the construction.   
Hence, even after the construction of the di-ring by moving the rings close to each other, there may still remain some mathematical structure (for example, a certain kind of symmetry) to ensure this fact. 
It may be also interesting whether this fact is realized only in the di-ring. 
%

%
As one of the further studies of the black di-ring, in Sec. III we analyzed a certain kind of stability of the di-ring from the thermodynamic viewpoint, which was introduced by Evslin and Krishnan for the study of black Saturn.  
Within the numeric search, we confirmed that the Hessian has at least one positive eigenvalue. 
This means that the thermodynamic black di-ring has no local stable state (metastable state). 
Furthermore, we found that in the narrow window of the thin branch near a cusp another eigenvalue also becomes positive. 
That is, we can say that in this region instability of the thermodynamic di-ring is enhanced.
This fact makes a remarkable contrast to the black Saturn, which has metastable states in the similar window. 
%

%
Finally we consider the behavior of the eigenvalue 2 near the maximal point as depicted in Fig. \ref{fig:eigen2_up}.
The eigenvalue 2 approaches zero promptly as $j^2 \to 1$, so that the corresponding mode seems to behave approximately like a zero mode. 
In the case of the thermodynamic black Saturn, near the extremal point we can show that the maximal eigenvalue of the Hessian has similar behavior to the eigenvalue 2 of the di-ring. 
Hence, together with the fact pointed out in Sec. III that the curves of the black ring, Saturn, and di-ring approach that of MP black hole, it may be suggested that near the extremal point, the MP black hole has large fluctuation modes corresponding to the approximately zero eigenvalues of the black objects.

\acknowledgments
We are grateful to Roberto Emapran and Pau Figueras for their generous proposal to submit their paper and the author's simultaneously to respect mutual priority. 
This work is partially supported by Grant-in-Aid for Young Scientists (B)
(No. 20740143) from the Japanese Ministry of Education, Science,
Sports, and Culture.

\clearpage

\appendix

\section{Physical variables of di-ring I}
\label{sec:variable1}
Using the exact expressions of black di-ring I, we can compute physical variables of the black di-ring.
The ADM mass $M$ and angular momentum $J$ are
\begin{equation}
M=\frac{3 \pi }{4}\left(a_{65}+a_{41}\frac{b_{\rm R}^2-b_{\rm L}^2+2 a_{41}}{(b_{\rm R}-b_{\rm L})^2}\right),
\end{equation}
and
\begin{equation}
J=\frac{\pi }{2}\frac{a_{41}}{(b_{\rm L}-b_{\rm R})^3}\left(4a_{41}^2-3(b_{\rm L}^2-b_{\rm R}^2)a_{41} + (b_{\rm L}-b_{\rm R})^2(b_{\rm L} b_{\rm R} +a_{71}+ a_{74} +  2 (a_{65}-a_{32}))\right).
\end{equation}
The Komar masses of outer and inner horizons are
\begin{equation}
M_{\rm K,L}=\frac{3\pi }{4}\frac{a_4-a_1}{b_{\rm R}-b_{\rm L}}\left(-\frac{(a_{54}a_{71}-a_{51}a_{74})a_{65}b_{\rm L} b_{\rm R}}{(a_{54}a_{61}a_{71}b_{\rm R}-a_{51}a_{64}a_{74}b_{\rm L})}+\frac{b_{\rm R}^2+2a_{41}-b_{\rm L}^2}{b_{\rm R}-b_{\rm L}}\right)
\end{equation}
and
\begin{equation}
M_{\rm K,R}=\frac{3 a_{65} (a_{61}b_{\rm R}-a_{64}b_{\rm L}) ( a_{54} a_{71}b_{\rm R}-a_{51}a_{74} b_{\rm L}) \pi }{4 (b_{\rm R}-b_{\rm L}) (a_{54} a_{61}a_{71} b_{\rm R}-a_{51} a_{64} a_{74}b_{\rm L})}.
\end{equation}
The Komar angular momenta of outer and inner horizons are
\begin{equation}
J_{\rm K,L}=\frac{2}{3 \Omega_{\rm L}}\left(M_{\rm K,L}-\frac{3\pi }{4}a_{32}\right),
\end{equation}
and
\begin{equation}
J_{\rm K,R}=\frac{2}{3 \Omega_{\rm R}}\left(M_{\rm K,R}-\frac{3\pi }{4}a_{65}\right).
\end{equation}
The angular velocities of outer and inner horizons are
\begin{equation}
\Omega_{\rm L}=\frac{(a_{42}b_{\rm L}+a_{21}b_{\rm R})(b_{\rm L}-b_{\rm R})}{a_{41}(2(a_{42}b_{\rm L}+a_{21}b_{\rm R})-b_{\rm L} b_{\rm R} (b_{\rm L}-b_{\rm R}))},
\end{equation}
and
\begin{equation}
\Omega_{\rm R}=\frac{a_{41}b_{\rm L} b_{\rm R} (b_{\rm L}-b_{\rm R})}{2\left((a_{64}a_{74}b_{\rm L}-a_{61}a_{71}b_{\rm R})(b_{\rm L}-b_{\rm R})+a_{41}^2b_{\rm L} b_{\rm R}\right)}.
\end{equation}
The areas and temperatures of horizons are
\begin{equation}
A_{h,\rm L} = 4\pi^2 a_{32} \sqrt{S_{\rm L}},
\end{equation}
\begin{equation}
A_{h,\rm R} = 4\pi^2 a_{65} \sqrt{S_{\rm R}},
\end{equation}
and
\begin{equation}
T_{\rm L}=\frac{1}{2\pi \sqrt{S_{\rm L}}},
\end{equation}
\begin{equation}
T_{\rm R}=\frac{1}{2\pi \sqrt{S_{\rm R}}},
\end{equation}
where
\begin{equation}
S_{\rm L}=4\frac{a_{32}^2a_{41}^2a_{62}(a_{64}a_{74}(a_{65}a_{71}+a_{73}a_{51})b_{\rm L}-a_{61}a_{71}(a_{65}a_{74}+a_{73}a_{54})b_{\rm R})^2}{a_{31}a_{43}a_{51}a_{52}a_{54}a_{61}a_{64}a_{71}a_{72}^2a_{74}(b_{\rm L}-b_{\rm R})^4},
\end{equation}
and
\begin{equation}
S_{\rm R}=4\frac{a_{65}^2a_{73}^2a_{62}(a_{42}a_{31}a_{51}(a_{61}a_{74}+a_{71}a_{64})b_{\rm L}-a_{61}a_{71}(a_{42}a_{31}a_{51}+a_{21}a_{43}a_{54})b_{\rm R})^2}{a_{42}a_{43}a_{54}a_{61}a_{63}a_{71}a_{72}^2a_{75}a_{31}^2a_{51}^2(b_{\rm L}-b_{\rm R})^4}.
\end{equation}
The periods $\Delta \phi$ for the spacelike rod $d_2$ and $d_4$ are given by
\begin{equation}
\left( \frac{\Delta \phi_{\rm L}}{2\pi} \right)^2=\frac{a_{53}a_{62}a_{73}^2(a_{42}a_{31}b_{\rm L}-a_{21}a_{43}b_{\rm R})^2}{a_{21}a_{31}a_{42}a_{43}a_{52}a_{63}a_{72}^2(b_{\rm L}-b_{\rm R})^2},
\end{equation}
and
\begin{equation}
\left( \frac{\Delta \phi_{\rm R}}{2\pi} \right)^2=\frac{a_{73}^2a_{76}(a_{74}b_{\rm L}-a_{71}b_{\rm R})^2}{a_{71}a_{72}^2a_{74}a_{75}(b_{\rm L}-b_{\rm R})^2}.
\end{equation}


\section{Physical variables of di-ring II}
\label{sec:variable2}
Using the exact expressions of black di-ring II, we can compute physical variables of the black di-ring.
The ADM mass $M$ and angular momentum $J$ are
\begin{equation}
 M = \frac{3 \pi}{4} (a_{31}+a_{64}),
\end{equation}
and
\begin{equation}
 J = - \frac{\pi}{2}  \frac{a_{21}a_{51}b_{\rm L} + a_{42}a_{54} b_{\rm R}}{a_{41}} .
\end{equation}
The Komar masses of outer and inner horizons are
\begin{equation}
 M_{\rm K,L} = \frac{3 \pi}{4}  \frac{a_{31}a_{43}(a_{51}b_{\rm L}-a_{54}b_{\rm R})}{a_{43}a_{51}b_{\rm L} + a_{31}a_{54}b_{\rm R}} ,
\end{equation}
and
\begin{equation}
 M_{\rm K,R} = \frac{3 \pi}{4} \frac{a_{43}a_{51}a_{64}b_{\rm L}+a_{31}a_{54}a_{61}b_{\rm R}}{a_{43}a_{51}b_{\rm L}+a_{31}a_{54}b_{\rm R}} .
\end{equation}
The Komar angular momenta of outer and inner horizons are
\begin{equation}
 J_{\rm K,L} = - \frac{\pi}{2} \frac{(a_{51}b_{\rm L}-a_{54}b_{\rm R})(a_{21}a_{43}a_{51}b_{\rm L}-a_{31}a_{42}a_{54}b_{\rm R})}
               {a_{41}(a_{43}a_{51}b_{\rm L}+a_{31}a_{54}b_{\rm R})} ,
\end{equation}
and
\begin{equation}
 J_{\rm K,R} = \frac{\pi}{2}  \frac{a_{41}a_{51}a_{54}b_{\rm L} b_{\rm R}}{a_{43}a_{51}b_{\rm L}+a_{31}a_{54}b_{\rm R}} .
\end{equation}
The angular velocities of outer and inner horizons are
\begin{equation}
 \Omega_{\rm L} = - \frac{a_{41}}{a_{51}b_{\rm L} - a_{54}b_{\rm R}},
\end{equation}
and
\begin{equation}
 \Omega_{\rm R} = - \frac{1}{a_{41}}\left( \frac{a_{31}}{b_{\rm L}} + \frac{a_{43}}{b_{\rm R}} \right).
\end{equation}
The areas and temperatures of horizons are
\begin{equation}
 A_{h,\rm L} = 4\pi^2 a_{32} \sqrt{S_{\rm L}},
\end{equation}
\begin{equation}
 A_{h,\rm R} = 4\pi^2 a_{65} \sqrt{S_{\rm R}}
\end{equation}
and
\begin{equation}
 T_{\rm L}=\frac{1}{2\pi \sqrt{S_{\rm L}}},
\end{equation}
\begin{equation}
 T_{\rm R}=\frac{1}{2\pi \sqrt{S_{\rm R}}},
\end{equation}
where
\begin{equation}
 S_{\rm L}=\frac{a_{21}a_{32}a_{42}a_{62}(a_{51}b_{\rm L}-a_{54}b_{\rm R})^2}{(a_{41}a_{52}a_{72})^2},
\end{equation}
and
\begin{equation}
 S_{\rm R}=\frac{2a_{61}a_{62}a_{64}a_{65}a_{71}a_{73}a_{74}}{(a_{63}a_{72}a_{75})^2}.
\end{equation}
The periods $\Delta \phi$ for the spacelike rod $d_2$ and $d_4$ are given by
\begin{equation}
 \left( \frac{\Delta \phi_{\rm L}}{2 \pi} \right)^2 =  \frac{a_{21}a_{42}a_{53}a_{62}a_{73}}{2a_{31}a_{43}} \left( \frac{a_{43}a_{51}b_{\rm L}+a_{31}a_{54}b_{\rm R}}{a_{41}a_{52}a_{63}a_{72}} \right)^2,
\end{equation}
and
\begin{equation}
 \left( \frac{\Delta \phi_{\rm R}}{2 \pi} \right)^2 = \frac{a_{71}a_{73}a_{74}a_{76}}{a_{72}{}^2 a_{75}{}^2}.
\end{equation}


\section{Physical variables of thermodynamic di-ring}
\label{sec:variable3}

Using the expressions of Appendix \ref{sec:variable1} or \ref{sec:variable2} with Eq. (\ref{eq:III7}) or (\ref{eq:III8}), respectively, we can reduce the physical variables of the black di-ring to the following simple forms under the constraints (\ref{eq:III5}) or (\ref{eq:III6}). 
For convenience, the rod length $d$ of the rings is introduced ($d=h_{1}-h_{2}=h_{3}-h_{4}$). 
The ADM mass $M$ and square of angular momentum $J^2$ are
\begin{equation}
M = \frac{3 \pi}{4} \frac{d(h_{1}+h_{4})^2}{h_{2}h_{4}},
\end{equation}
and
\begin{equation}
 J^2 =  \frac{\pi^2}{2} \frac{d^2h_{1}h_{3}(h_{1}+h_{4})(h_{1}^2+(d+h_{3})h_{4})^2}{h^{3}_{2}h^{3}_{4}}.
\end{equation}
The Komar masses of outer and inner horizons are
\begin{equation}
 M_{\rm K,L} = \frac{3 \pi}{4} \frac{d h_{1}(h_{1}+h_{4})}{h_{2}h_{4}},
\end{equation}
and
\begin{equation}
M_{\rm K,R} = \frac{3 \pi}{4} \frac{d(h_{1}+h_{4})}{h_{2}}.
\end{equation}
The squares of Komar angular momenta of outer and inner horizons are
\begin{equation}
 J^{2}_{\rm K,L} = \frac{\pi^2}{2} \frac{d^2h_{1}h_{3}(h_{1}+h_{4})(h_{1}^2+d h_{4})^2}{h^{3}_{2}h^{3}_{4}},
\end{equation}
and 
\begin{equation}
 J^{2}_{\rm K,R} = \frac{\pi^2}{2} \frac{d^2h_{1}h^{3}_{3}(h_{1}+h_{4})}{h^{3}_{2}h_{4}}.
\end{equation}
The squares of angular velocities of outer and inner horizons are 
\begin{equation}
 \Omega_{\rm L}^{2} = \Omega_{\rm R}^{2} = \frac{h_{2}h_{4}}{2h_{1}h_{3}(h_{1}+h_{4})}.
\end{equation}
The areas and temperatures of horizons are
\begin{equation}
 A_{h,\rm L} = A_{h,\rm R} = 4\pi^2 d \sqrt{S}
\end{equation}
and
\begin{equation}
T_{h,\rm L} = T_{h,\rm R} = \frac{1}{2\pi \sqrt{S}},
\end{equation}
where
\begin{equation}
S=\frac{2 d^2 h_{3}(h_{1}+h_{4})(h_{1}-h_{4})^2}{(h_{1}-h_{3})^2h_{2}h^{2}_{4}}.
\end{equation}

\bibliography{TDR0819_1.bib}

\begin{thebibliography}{36}
\expandafter\ifx\csname natexlab\endcsname\relax\def\natexlab#1{#1}\fi
\expandafter\ifx\csname bibnamefont\endcsname\relax
  \def\bibnamefont#1{#1}\fi
\expandafter\ifx\csname bibfnamefont\endcsname\relax
  \def\bibfnamefont#1{#1}\fi
\expandafter\ifx\csname citenamefont\endcsname\relax
  \def\citenamefont#1{#1}\fi
\expandafter\ifx\csname url\endcsname\relax
  \def\url#1{\texttt{#1}}\fi
\expandafter\ifx\csname urlprefix\endcsname\relax\def\urlprefix{URL }\fi
\providecommand{\bibinfo}[2]{#2}
\providecommand{\eprint}[2][]{\url{#2}}

\bibitem[{\citenamefont{Myers and Perry}(1986)}]{Myers:1986un}
\bibinfo{author}{\bibfnamefont{R.~C.} \bibnamefont{Myers}} \bibnamefont{and}
  \bibinfo{author}{\bibfnamefont{M.~J.} \bibnamefont{Perry}},
  \bibinfo{journal}{Ann. Phys.} \textbf{\bibinfo{volume}{172}},
  \bibinfo{pages}{304} (\bibinfo{year}{1986}).

\bibitem[{\citenamefont{Emparan and Reall}(2002)}]{Emparan:2001wn}
\bibinfo{author}{\bibfnamefont{R.}~\bibnamefont{Emparan}} \bibnamefont{and}
  \bibinfo{author}{\bibfnamefont{H.~S.} \bibnamefont{Reall}},
  \bibinfo{journal}{Phys. Rev. Lett.} \textbf{\bibinfo{volume}{88}},
  \bibinfo{pages}{101101} (\bibinfo{year}{2002}), \eprint{hep-th/0110260}.

\bibitem[{\citenamefont{Pomeransky}(2006)}]{Pomeransky:2005sj}
\bibinfo{author}{\bibfnamefont{A.~A.} \bibnamefont{Pomeransky}},
  \bibinfo{journal}{Phys. Rev.} \textbf{\bibinfo{volume}{D73}},
  \bibinfo{pages}{044004} (\bibinfo{year}{2006}), \eprint{hep-th/0507250}.

\bibitem[{\citenamefont{Iguchi and
  Mishima}(2006{\natexlab{a}})}]{Iguchi:2006rd}
\bibinfo{author}{\bibfnamefont{H.}~\bibnamefont{Iguchi}} \bibnamefont{and}
  \bibinfo{author}{\bibfnamefont{T.}~\bibnamefont{Mishima}},
  \bibinfo{journal}{Phys. Rev.} \textbf{\bibinfo{volume}{D73}},
  \bibinfo{pages}{121501} (\bibinfo{year}{2006}{\natexlab{a}}),
  \eprint{hep-th/0604050}.

\bibitem[{\citenamefont{Tomizawa and Nozawa}(2006)}]{Tomizawa:2006vp}
\bibinfo{author}{\bibfnamefont{S.}~\bibnamefont{Tomizawa}} \bibnamefont{and}
  \bibinfo{author}{\bibfnamefont{M.}~\bibnamefont{Nozawa}},
  \bibinfo{journal}{Phys. Rev.} \textbf{\bibinfo{volume}{D73}},
  \bibinfo{pages}{124034} (\bibinfo{year}{2006}), \eprint{hep-th/0604067}.

\bibitem[{\citenamefont{Mishima and Iguchi}(2006)}]{Mishima:2005id}
\bibinfo{author}{\bibfnamefont{T.}~\bibnamefont{Mishima}} \bibnamefont{and}
  \bibinfo{author}{\bibfnamefont{H.}~\bibnamefont{Iguchi}},
  \bibinfo{journal}{Phys. Rev.} \textbf{\bibinfo{volume}{D73}},
  \bibinfo{pages}{044030} (\bibinfo{year}{2006}), \eprint{hep-th/0504018}.

\bibitem[{\citenamefont{Iguchi and
  Mishima}(2006{\natexlab{b}})}]{Iguchi:2006tu}
\bibinfo{author}{\bibfnamefont{H.}~\bibnamefont{Iguchi}} \bibnamefont{and}
  \bibinfo{author}{\bibfnamefont{T.}~\bibnamefont{Mishima}},
  \bibinfo{journal}{Phys. Rev.} \textbf{\bibinfo{volume}{D74}},
  \bibinfo{pages}{024029} (\bibinfo{year}{2006}{\natexlab{b}}),
  \eprint{hep-th/0605090}.

\bibitem[{\citenamefont{Tomizawa
  et~al.}(2006{\natexlab{a}})\citenamefont{Tomizawa, Morisawa, and
  Yasui}}]{Tomizawa:2005wv}
\bibinfo{author}{\bibfnamefont{S.}~\bibnamefont{Tomizawa}},
  \bibinfo{author}{\bibfnamefont{Y.}~\bibnamefont{Morisawa}}, \bibnamefont{and}
  \bibinfo{author}{\bibfnamefont{Y.}~\bibnamefont{Yasui}},
  \bibinfo{journal}{Phys. Rev.} \textbf{\bibinfo{volume}{D73}},
  \bibinfo{pages}{064009} (\bibinfo{year}{2006}{\natexlab{a}}),
  \eprint{hep-th/0512252}.

\bibitem[{\citenamefont{Pomeransky and Sen'kov}(2006)}]{Pomeransky:2006bd}
\bibinfo{author}{\bibfnamefont{A.~A.} \bibnamefont{Pomeransky}}
  \bibnamefont{and} \bibinfo{author}{\bibfnamefont{R.~A.}
  \bibnamefont{Sen'kov}} (\bibinfo{year}{2006}), \eprint{hep-th/0612005}.

\bibitem[{\citenamefont{Elvang and Figueras}(2007)}]{Elvang:2007rd}
\bibinfo{author}{\bibfnamefont{H.}~\bibnamefont{Elvang}} \bibnamefont{and}
  \bibinfo{author}{\bibfnamefont{P.}~\bibnamefont{Figueras}},
  \bibinfo{journal}{JHEP} \textbf{\bibinfo{volume}{05}}, \bibinfo{pages}{050}
  (\bibinfo{year}{2007}), \eprint{hep-th/0701035}.

\bibitem[{\citenamefont{Iguchi and Mishima}(2007)}]{Iguchi:2007is}
\bibinfo{author}{\bibfnamefont{H.}~\bibnamefont{Iguchi}} \bibnamefont{and}
  \bibinfo{author}{\bibfnamefont{T.}~\bibnamefont{Mishima}},
  \bibinfo{journal}{Phys. Rev.} \textbf{\bibinfo{volume}{D75}},
  \bibinfo{pages}{064018} (\bibinfo{year}{2007}), \eprint{hep-th/0701043}.

\bibitem[{\citenamefont{Evslin and Krishnan}(2009)}]{Evslin:2007fv}
\bibinfo{author}{\bibfnamefont{J.}~\bibnamefont{Evslin}} \bibnamefont{and}
  \bibinfo{author}{\bibfnamefont{C.}~\bibnamefont{Krishnan}},
  \bibinfo{journal}{Class. Quant. Grav.} \textbf{\bibinfo{volume}{26}},
  \bibinfo{pages}{125018} (\bibinfo{year}{2009}), \eprint{0706.1231}.

\bibitem[{\citenamefont{Izumi}(2008)}]{Izumi:2007qx}
\bibinfo{author}{\bibfnamefont{K.}~\bibnamefont{Izumi}},
  \bibinfo{journal}{Prog. Theor. Phys.} \textbf{\bibinfo{volume}{119}},
  \bibinfo{pages}{757} (\bibinfo{year}{2008}), \eprint{0712.0902}.

\bibitem[{\citenamefont{Elvang and Rodriguez}(2008)}]{Elvang:2007hs}
\bibinfo{author}{\bibfnamefont{H.}~\bibnamefont{Elvang}} \bibnamefont{and}
  \bibinfo{author}{\bibfnamefont{M.~J.} \bibnamefont{Rodriguez}},
  \bibinfo{journal}{JHEP} \textbf{\bibinfo{volume}{04}}, \bibinfo{pages}{045}
  (\bibinfo{year}{2008}), \eprint{0712.2425}.

\bibitem[{\citenamefont{Evslin}(2008)}]{Evslin:2008gx}
\bibinfo{author}{\bibfnamefont{J.}~\bibnamefont{Evslin}},
  \bibinfo{journal}{JHEP} \textbf{\bibinfo{volume}{09}}, \bibinfo{pages}{004}
  (\bibinfo{year}{2008}), \eprint{0806.3389}.

\bibitem[{\citenamefont{Chen and Teo}(2008)}]{Chen:2008fa}
\bibinfo{author}{\bibfnamefont{Y.}~\bibnamefont{Chen}} \bibnamefont{and}
  \bibinfo{author}{\bibfnamefont{E.}~\bibnamefont{Teo}},
  \bibinfo{journal}{Phys. Rev.} \textbf{\bibinfo{volume}{D78}},
  \bibinfo{pages}{064062} (\bibinfo{year}{2008}), \eprint{0808.0587}.

\bibitem[{\citenamefont{Dias et~al.}(2009)\citenamefont{Dias, Figueras,
  Monteiro, Santos, and Emparan}}]{Dias:2009iu}
\bibinfo{author}{\bibfnamefont{O.~J.~C.} \bibnamefont{Dias}},
  \bibinfo{author}{\bibfnamefont{P.}~\bibnamefont{Figueras}},
  \bibinfo{author}{\bibfnamefont{R.}~\bibnamefont{Monteiro}},
  \bibinfo{author}{\bibfnamefont{J.~E.} \bibnamefont{Santos}},
  \bibnamefont{and} \bibinfo{author}{\bibfnamefont{R.}~\bibnamefont{Emparan}},
  \bibinfo{journal}{Phys. Rev.} \textbf{\bibinfo{volume}{D80}},
  \bibinfo{pages}{111701} (\bibinfo{year}{2009}), \eprint{0907.2248}.

\bibitem[{\citenamefont{Dias et~al.}(2010{\natexlab{a}})\citenamefont{Dias,
  Figueras, Monteiro, Reall, and Santos}}]{Dias:2010eu}
\bibinfo{author}{\bibfnamefont{O.~J.~C.} \bibnamefont{Dias}},
  \bibinfo{author}{\bibfnamefont{P.}~\bibnamefont{Figueras}},
  \bibinfo{author}{\bibfnamefont{R.}~\bibnamefont{Monteiro}},
  \bibinfo{author}{\bibfnamefont{H.~S.} \bibnamefont{Reall}}, \bibnamefont{and}
  \bibinfo{author}{\bibfnamefont{J.~E.} \bibnamefont{Santos}},
  \bibinfo{journal}{JHEP} \textbf{\bibinfo{volume}{05}}, \bibinfo{pages}{076}
  (\bibinfo{year}{2010}{\natexlab{a}}), \eprint{1001.4527}.

\bibitem[{\citenamefont{Dias et~al.}(2010{\natexlab{b}})\citenamefont{Dias,
  Figueras, Monteiro, and Santos}}]{Dias:2010maa}
\bibinfo{author}{\bibfnamefont{O.~J.~C.} \bibnamefont{Dias}},
  \bibinfo{author}{\bibfnamefont{P.}~\bibnamefont{Figueras}},
  \bibinfo{author}{\bibfnamefont{R.}~\bibnamefont{Monteiro}}, \bibnamefont{and}
  \bibinfo{author}{\bibfnamefont{J.~E.} \bibnamefont{Santos}}
  (\bibinfo{year}{2010}{\natexlab{b}}), \eprint{1006.1904}.

\bibitem[{\citenamefont{Emparan et~al.}(2010)\citenamefont{Emparan, Harmark,
  Niarchos, and Obers}}]{Emparan:2009vd}
\bibinfo{author}{\bibfnamefont{R.}~\bibnamefont{Emparan}},
  \bibinfo{author}{\bibfnamefont{T.}~\bibnamefont{Harmark}},
  \bibinfo{author}{\bibfnamefont{V.}~\bibnamefont{Niarchos}}, \bibnamefont{and}
  \bibinfo{author}{\bibfnamefont{N.~A.} \bibnamefont{Obers}},
  \bibinfo{journal}{JHEP} \textbf{\bibinfo{volume}{04}}, \bibinfo{pages}{046}
  (\bibinfo{year}{2010}), \eprint{0912.2352}.

\bibitem[{\citenamefont{Shibata and Yoshino}(2010)}]{Shibata:2010wz}
\bibinfo{author}{\bibfnamefont{M.}~\bibnamefont{Shibata}} \bibnamefont{and}
  \bibinfo{author}{\bibfnamefont{H.}~\bibnamefont{Yoshino}},
  \bibinfo{journal}{Phys. Rev.} \textbf{\bibinfo{volume}{D81}},
  \bibinfo{pages}{104035} (\bibinfo{year}{2010}), \eprint{1004.4970}.

\bibitem[{\citenamefont{Elvang et~al.}(2007)\citenamefont{Elvang, Emparan, and
  Figueras}}]{Elvang:2007hg}
\bibinfo{author}{\bibfnamefont{H.}~\bibnamefont{Elvang}},
  \bibinfo{author}{\bibfnamefont{R.}~\bibnamefont{Emparan}}, \bibnamefont{and}
  \bibinfo{author}{\bibfnamefont{P.}~\bibnamefont{Figueras}},
  \bibinfo{journal}{JHEP} \textbf{\bibinfo{volume}{05}}, \bibinfo{pages}{056}
  (\bibinfo{year}{2007}), \eprint{hep-th/0702111}.

\bibitem[{\citenamefont{Emparan et~al.}(2007)\citenamefont{Emparan, Harmark,
  Niarchos, Obers, and Rodriguez}}]{Emparan:2007wm}
\bibinfo{author}{\bibfnamefont{R.}~\bibnamefont{Emparan}},
  \bibinfo{author}{\bibfnamefont{T.}~\bibnamefont{Harmark}},
  \bibinfo{author}{\bibfnamefont{V.}~\bibnamefont{Niarchos}},
  \bibinfo{author}{\bibfnamefont{N.~A.} \bibnamefont{Obers}}, \bibnamefont{and}
  \bibinfo{author}{\bibfnamefont{M.~J.} \bibnamefont{Rodriguez}},
  \bibinfo{journal}{JHEP} \textbf{\bibinfo{volume}{10}}, \bibinfo{pages}{110}
  (\bibinfo{year}{2007}), \eprint{0708.2181}.

\bibitem[{\citenamefont{Hollands and Yazadjiev}(2008)}]{Hollands:2007aj}
\bibinfo{author}{\bibfnamefont{S.}~\bibnamefont{Hollands}} \bibnamefont{and}
  \bibinfo{author}{\bibfnamefont{S.}~\bibnamefont{Yazadjiev}},
  \bibinfo{journal}{Commun. Math. Phys.} \textbf{\bibinfo{volume}{283}},
  \bibinfo{pages}{749} (\bibinfo{year}{2008}), \eprint{0707.2775}.

\bibitem[{\citenamefont{Tomizawa
  et~al.}(2006{\natexlab{b}})\citenamefont{Tomizawa, Iguchi, and
  Mishima}}]{Tomizawa:2006jz}
\bibinfo{author}{\bibfnamefont{S.}~\bibnamefont{Tomizawa}},
  \bibinfo{author}{\bibfnamefont{H.}~\bibnamefont{Iguchi}}, \bibnamefont{and}
  \bibinfo{author}{\bibfnamefont{T.}~\bibnamefont{Mishima}},
  \bibinfo{journal}{Phys. Rev.} \textbf{\bibinfo{volume}{D74}},
  \bibinfo{pages}{104004} (\bibinfo{year}{2006}{\natexlab{b}}),
  \eprint{hep-th/0608169}.

\bibitem[{\citenamefont{Belinsky and Zakharov}(1978)}]{Belinsky:1971nt}
\bibinfo{author}{\bibfnamefont{V.~A.} \bibnamefont{Belinsky}} \bibnamefont{and}
  \bibinfo{author}{\bibfnamefont{V.~E.} \bibnamefont{Zakharov}},
  \bibinfo{journal}{Sov. Phys. JETP} \textbf{\bibinfo{volume}{48}},
  \bibinfo{pages}{985} (\bibinfo{year}{1978}).

\bibitem[{\citenamefont{Belinsky and Sakharov}(1979)}]{Belinsky:1979mh}
\bibinfo{author}{\bibfnamefont{V.~A.} \bibnamefont{Belinsky}} \bibnamefont{and}
  \bibinfo{author}{\bibfnamefont{V.~E.} \bibnamefont{Sakharov}},
  \bibinfo{journal}{Sov. Phys. JETP} \textbf{\bibinfo{volume}{50}},
  \bibinfo{pages}{1} (\bibinfo{year}{1979}).

\bibitem[{\citenamefont{Harmark}(2004)}]{Harmark:2004rm}
\bibinfo{author}{\bibfnamefont{T.}~\bibnamefont{Harmark}},
  \bibinfo{journal}{Phys. Rev.} \textbf{\bibinfo{volume}{D70}},
  \bibinfo{pages}{124002} (\bibinfo{year}{2004}), \eprint{hep-th/0408141}.

\bibitem[{\citenamefont{Morisawa and Ida}(2004)}]{Morisawa:2004tc}
\bibinfo{author}{\bibfnamefont{Y.}~\bibnamefont{Morisawa}} \bibnamefont{and}
  \bibinfo{author}{\bibfnamefont{D.}~\bibnamefont{Ida}},
  \bibinfo{journal}{Phys. Rev.} \textbf{\bibinfo{volume}{D69}},
  \bibinfo{pages}{124005} (\bibinfo{year}{2004}), \eprint{gr-qc/0401100}.

\bibitem[{\citenamefont{Morisawa et~al.}(2008)\citenamefont{Morisawa, Tomizawa,
  and Yasui}}]{Morisawa:2007di}
\bibinfo{author}{\bibfnamefont{Y.}~\bibnamefont{Morisawa}},
  \bibinfo{author}{\bibfnamefont{S.}~\bibnamefont{Tomizawa}}, \bibnamefont{and}
  \bibinfo{author}{\bibfnamefont{Y.}~\bibnamefont{Yasui}},
  \bibinfo{journal}{Phys. Rev.} \textbf{\bibinfo{volume}{D77}},
  \bibinfo{pages}{064019} (\bibinfo{year}{2008}), \eprint{0710.4600}.

\bibitem[{\citenamefont{Armas and Harmark}(2010)}]{Armas:2009dd}
\bibinfo{author}{\bibfnamefont{J.}~\bibnamefont{Armas}} \bibnamefont{and}
  \bibinfo{author}{\bibfnamefont{T.}~\bibnamefont{Harmark}},
  \bibinfo{journal}{JHEP} \textbf{\bibinfo{volume}{05}}, \bibinfo{pages}{093}
  (\bibinfo{year}{2010}), \eprint{0911.4654}.

\bibitem[{\citenamefont{Elvang et~al.}(2006)\citenamefont{Elvang, Emparan, and
  Virmani}}]{Elvang:2006dd}
\bibinfo{author}{\bibfnamefont{H.}~\bibnamefont{Elvang}},
  \bibinfo{author}{\bibfnamefont{R.}~\bibnamefont{Emparan}}, \bibnamefont{and}
  \bibinfo{author}{\bibfnamefont{A.}~\bibnamefont{Virmani}},
  \bibinfo{journal}{JHEP} \textbf{\bibinfo{volume}{12}}, \bibinfo{pages}{074}
  (\bibinfo{year}{2006}), \eprint{hep-th/0608076}.

\bibitem[{\citenamefont{Evslin and Krishnan}(2008)}]{Evslin:2008py}
\bibinfo{author}{\bibfnamefont{J.}~\bibnamefont{Evslin}} \bibnamefont{and}
  \bibinfo{author}{\bibfnamefont{C.}~\bibnamefont{Krishnan}},
  \bibinfo{journal}{JHEP} \textbf{\bibinfo{volume}{09}}, \bibinfo{pages}{003}
  (\bibinfo{year}{2008}), \eprint{0804.4575}.

\bibitem[{\citenamefont{Herdeiro
  et~al.}(2010{\natexlab{a}})\citenamefont{Herdeiro, Kleihaus, Kunz, and
  Radu}}]{Herdeiro:2009vd}
\bibinfo{author}{\bibfnamefont{C.}~\bibnamefont{Herdeiro}},
  \bibinfo{author}{\bibfnamefont{B.}~\bibnamefont{Kleihaus}},
  \bibinfo{author}{\bibfnamefont{J.}~\bibnamefont{Kunz}}, \bibnamefont{and}
  \bibinfo{author}{\bibfnamefont{E.}~\bibnamefont{Radu}},
  \bibinfo{journal}{Phys. Rev.} \textbf{\bibinfo{volume}{D81}},
  \bibinfo{pages}{064013} (\bibinfo{year}{2010}{\natexlab{a}}),
  \eprint{0912.3386}.

\bibitem[{\citenamefont{Herdeiro
  et~al.}(2010{\natexlab{b}})\citenamefont{Herdeiro, Radu, and
  Rebelo}}]{Herdeiro:2010aq}
\bibinfo{author}{\bibfnamefont{C.}~\bibnamefont{Herdeiro}},
  \bibinfo{author}{\bibfnamefont{E.}~\bibnamefont{Radu}}, \bibnamefont{and}
  \bibinfo{author}{\bibfnamefont{C.}~\bibnamefont{Rebelo}},
  \bibinfo{journal}{Phys. Rev.} \textbf{\bibinfo{volume}{D81}},
  \bibinfo{pages}{104031} (\bibinfo{year}{2010}{\natexlab{b}}),
  \eprint{1004.3959}.

\bibitem[{\citenamefont{Astefanesei et~al.}(2009)\citenamefont{Astefanesei,
  Rodriguez, and Theisen}}]{Astefanesei:2009mc}
\bibinfo{author}{\bibfnamefont{D.}~\bibnamefont{Astefanesei}},
  \bibinfo{author}{\bibfnamefont{M.~J.} \bibnamefont{Rodriguez}},
  \bibnamefont{and} \bibinfo{author}{\bibfnamefont{S.}~\bibnamefont{Theisen}},
  \bibinfo{journal}{JHEP} \textbf{\bibinfo{volume}{12}}, \bibinfo{pages}{040}
  (\bibinfo{year}{2009}), \eprint{0909.0008}.

\end{thebibliography}
\end{document}